\documentclass[twoside]{article} \textheight 9in\textwidth 6.5in
 \usepackage{amssymb,epsfig} 
\begin{document}
 \newtheorem{dfn} {Definition} \newtheorem{rem} {Remark}
 \newtheorem{theorem}{Theorem} \newtheorem{lemma}{Lemma}
 \newtheorem{cor}{Corollary} \newcommand{\ie}{{\em i.e. }}
 \newcommand\eg{{\em e.g., }} \newcommand\qed{\rule{1em}{1.5ex}}
 \newcommand\fd[1]{{\em\bf #1}} \newcommand\df[1]{{\bf\em #1}}

\title {\vspace*{-3pc}Aperiodic Tilings: Breaking Translational Symmetry}

\author {Leonid A.~Levin\thanks {Supported by NSF grant CCR-0311411.}\\
 Boston University\thanks {Computer Science department, 111 Cummington
St., Boston, MA 02215; (email: Lnd at bu.edu)}} \date{}\maketitle

\begin{abstract} Classical results on aperiodic tilings are rather
complicated and not widely understood. Below, an alternative approach is
discussed in hope to provide additional intuition not apparent in
classical works. \end{abstract}

\section {Palettes and Tilings}

Physical computing media are asymmetric. Their symmetry is broken
by irregularities, physical boundaries, external connections, and so on.
 Such peculiarities, however, are highly variable and extraneous
 to the fundamental nature of the media. Thus, they are pruned
 from theoretical models, such as cellular automata, and
 reliance on them is frowned upon in programming practice.

Yet, computation, like many other highly organized activities, is
incompatible with perfect symmetry. Some standard mechanisms must assure
breaking the symmetry inherent in idealized computing models.
 A famous example of such mechanisms is aperiodic tiling: hierarchical
self-similar constructions, first used for computational purposes in a
classical -- although rather complicated -- work \cite{berger} and
further developed in \cite{robinson, myers,gurkor72}. \cite{ad-bgg} give
a helpful exposition.

\begin{dfn} Let $G$ be the grid of unit length edges between integer
points on an infinite plane.\par
 A \fd {tiling} $T$ is its mapping into a finite set of \fd {colors}.
Its \fd {crosses} and \fd {tiles} are ordered color combinations of four
edges sharing a corner or forming a square, respectively. A \fd{palette}
$P$ of $T$ is a set including all its tiles (\fd{+palette} for crosses).
 We say $P$ with a mapping $f$ of its colors into a smaller color
alphabet \fd {enforces} a set $S$ of tilings if replacing colors
according to $f$ turns each $P$-tiling into a tiling in $S$.\end{dfn}

Turning each edge orthogonally around its center turns $G$ into
its dual graph and palettes into +palettes and vice versa.
Thus, one can use either type as convenient.

\section {2-Adic Coordinates}

The set of all tilings with a given palette $P$ has translational
symmetry, \ie any shift produces another $P$-tiling. We want a palette
that forces a complete spontaneous breaking of this symmetry, \ie
prevents individual tilings from being {\em periodic}.
 Accordingly, each location in a given tiling will be uniquely characterized
by a sort of {\em coordinates}. Their infinitely many values cannot be
reflected in the finite variety of the tile's colors. They will be
represented {\em distributively}, \ie in the colors of the surrounding
tiles, and computable from them to any given number of digits.

Let us first so distribute the horizontal Cartesian integer coordinates
$x=(2i\!+\!1)2^k$ of vertical edges by reflecting one bit $(i\bmod 2)$
in their color. We view this bit as the direction of a \df {bracket}. In
this 1-dimensional tiling $C_1$, the brackets of the same \df {rank} $k$
are equidistant (Figure~\ref{C1-by-rank}).

\newcommand\ranks{\centering \fbox{\begin{minipage}{3in}{\begin{tabbing}
 [ \=\ \ ] \=\ \ [ \=\ \ ] \=\ \ [ \=\ \ ] \=\ \ [ \=\ \
 ] \=\ \ [ \=\ \ ] \=\ \ [ \=\ \ ] \=\ \ [ \=\ \ ]\+\\
 \{\>\>\}\>\>\{\>\>\}\>\>\{\>\>\}\>\>\{\+\\\+[\>\>\>\>]\>\>\>\>[\+
\\\{\>\>\>\>\>\>\>\>\}\+\+\+\\\>[ \end{tabbing}}\end{minipage}}
 \caption {Brackets of $C_1$ split by rank. \label {C1-by-rank}}}

It is convenient to visualize the bits of even rank, picturing
them \}/\{ or red, or dotted, separately from odd, depicted ]/[.
The bits of either shape at each side of the origin form a progression
of balanced parenthetical expressions, called \df {domains}.
Each domain has four \df {grandchildren} of the second lower rank:
two within its outermost brackets and one to each side.
The two \df {children} have the other shape and are centered at each
border of the \df {parent}, thus connecting it with its grandchildren.

\begin{figure}[ht]
 \hspace*{\fill}\parbox[c]{3in}{\ranks
 \caption{
 (Right)\hfill
 Boldness bit: bold lines in $C$.
 {Courtesy of A. Shen and B. Durand.} \label {Boldness-in-C}}
 } \hfill\parbox[c]{3in}
 {\includegraphics[scale=0.6]{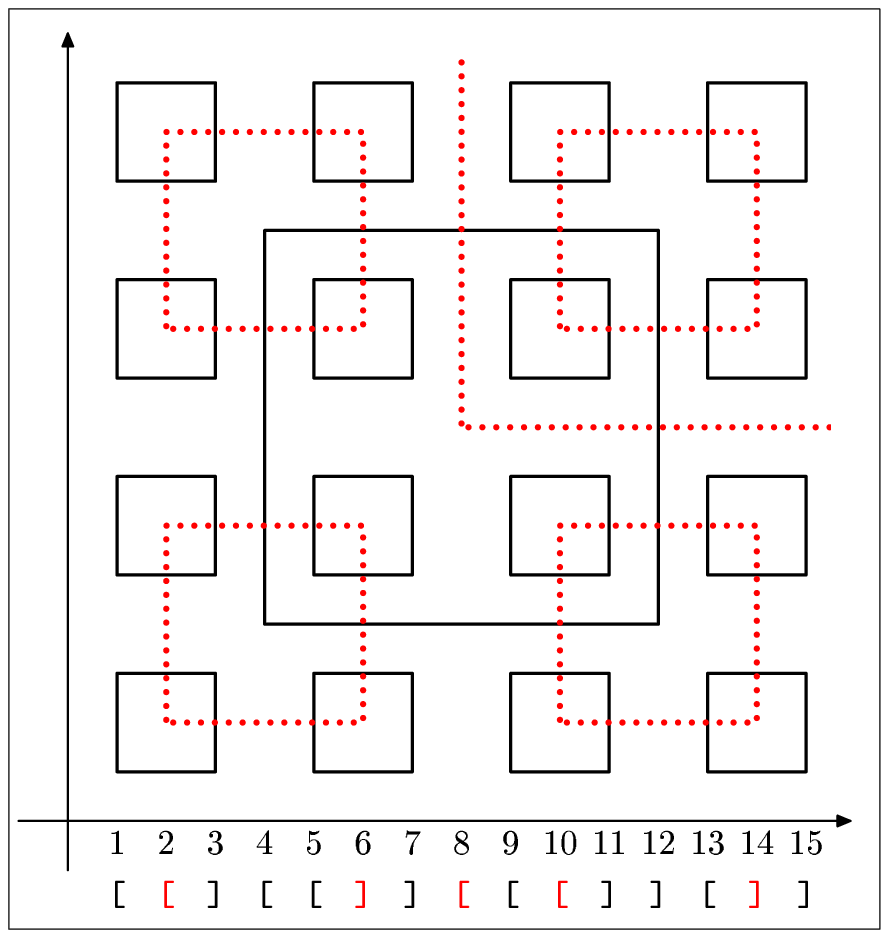}}
 \end{figure}

Handling the vertical coordinate similarly yields a neat 2-dimensional
tiling $C$ called \df {central}. Figure~\ref{Boldness-in-C} marks the
borders of intersections of its vertical and horizontal domains of equal
ranks with a \df {boldness} bit.

$C_1$ has a special, \ie unmatched, bracket in the origin, directed
arbitrarily and unranked. No palette can enforce a set of tilings with
unique special points (designated by a Borel function commuting with
shifts) since the set of all tilings is compact\footnote
 {and so has finite shift-invariant measures, \eg
  defined by condensation points of frequencies of
  finite configurations in some quasiperiodic tiling}
 whereas the set of locations of their special point and the group
$\mathbb{Z}$ of their shifts is not. We will extend $\mathbb{Z}$ to a
compact group and also \df {define ranks} in other tilings, \eg shifted
$C_1$, using the following property.

\begin{rem}\label{R} A shift by $(2i\!+\!1)2^k$ in $C_1$
 reverses all brackets of rank $k\!-\!1$, none of lower ranks,
 and every second bracket of any rank $>k$.\end{rem}

Therefore, the shifts by $(2i\!+\!1)2^k$ change our bits only at a $2/2^k$
fraction of locations. This fraction can be used as a metric on the
group of shifts which can then be completed for it. The result is a
remarkable compact group $g$ of 2-adic integers, or \df {2-adics}, acting
on a similarly completed set of 2-adic coordinates. A 2-adic $a$ is a
formal infinite sum $\sum_{i\ge0} 2^i a_i=\ldots+4a_2+2a_1+a_0$, where
$a_i\in\{0,1\}$, viewed as an infinite to the left sequence of bits.
 The usual algorithms for addition and multiplication make $g$ a ring
with $\mathbb{Z}$ as a subring (\eg $-1=\ldots+8+4+2+1$).\footnote
 {\samepage Odd 2-adics have inverses. This allows extending
  $g$ to a famous locally compact field with fractions $a/2^i$.}

The natural action of $\mathbb{Z}$ (by shifts) can be extended
to the action of the whole $g$ on $C_1$ and its images.
 Indeed, the brackets of rank $k$ are unaffected by terms $a_i$ with
$i>k\!+\!1$. Thus, a 2-adic shift $a$ of $C_1$ can be defined as the
pointwise limit of the sequence of shifts by integers approximating $a$.

With inverse shifts, this sequence diverges for the unranked bracket in
the origin of $C_1$ and of its integer shifts.
 The direction of this bracket is determined not by its location, but by
an arbitrary \df {default} included as an additional (external, unmoved
by shifts) point in the tilings. The \df {reflection} reverses the
default, all brackets, and the signs of their locations.
 With added reflection, the action of $g$ is free and transitive:
 each of these tilings can be obtained from any other, \eg from $C_1$.
 In 2 dimensions we can also add the diagonal reflection
exchanging the vertical and horizontal coordinates.\footnote
   {We allow fewer tilings than \cite{ad-bgg} which
    permits different shifts at each side of the origin.}

 \section {Enforcing the Coordinate System with a Palette}

\begin{theorem}\label{E} The set of 2-adic shifts and reflections
of tiling $C$ can be enforced by a palette. \end{theorem}

To prove it, we use multicomponent colors in $T$ which $f$ projects
onto their first component -- bracket bit. The second component includes
two \df {enforcement} bits that extend $C$ to the enhanced tiling $CE$,
with a +palette $ce$ of 7 crosses modulo 8 reflections.\footnote
 {\cite{robinson} uses only six tiles (with reflections)
  but colors their corners, in addition to sides.}
 One of its bits is the already described boldness bit (Figure~\ref
{Boldness-in-C}). The other is a \df {pointer} in the direction of the
nearest orthogonal line of the same rank. On the (unranked) axes these
bits are set by a default central cross.

\begin{dfn} A \fd {box} is an \fd {open} $ce$-tiled rectangle,
 \ie one with the border edges removed.
 Its \fd {$k$-block} is a square with monochromatic sides that is\footnote
 {The rest of the requirement is redundant but useful in the proof.}
 a tile (for $k\!=\!0$) or a combination of four $(k\!-\!1)$-blocks
sharing a corner. The four segments connecting the block's center to its
sides are called \fd {$(k\!\!-\!\!1) $-medians}.
 The rest of the open block is called a \fd {frame}.
 We call a box \fd {$k$-tiled} if removing an outer layer which is
thinner than a $k$-block turns it into a box composed of (open at the
box border) $k$-blocks.\footnote {$ce$ prevents crossing of monochromatic
segments, making decompositions of boxes into blocks unique.}\end{dfn}

\begin{lemma}\label{blocks}
(i) Borders between open blocks in a box are monochromatic.
(ii) All $k$-frame patterns in a box are enclosed in its open $k$-blocks.
(iii) All open $k$-blocks are congruent and have equal frames.\end{lemma}

\paragraph {Proof.} (i) comes from all $ce$-crosses having one or four
inward pointers. (ii,iii) for $k\!>\!2$ follow from $k\!-\!1$ by viewing
1-blocks as tiles. Let 1-blocks $a$ and $b$ be adjacent in a 2-block
$c$; $(L,l)$ and $(R,r)$ be pairs of medians of $a$ and $b$ with $l,r$
directed to the side $s$ of $c$, $L\!-\!R$ crossing a median $m$ of $c$
at a cross $x$. $x$ forces $L,R$ to be both pale or both bold. This
forces opposite brackets on $l,r$ which, too, must be both pale or both
bold depending on the bracket of $L\!-\!R$. $l,r$ cannot be both bold
which would require the pointer of $s$ to agree with their opposite
brackets. Thus, all external medians of 2-frames are pale, internal
medians bold, their brackets face the frame's center forcing inward
pointers on the 1-medians, like $m$. \qed

\begin{lemma}\label{I} Any 1-tiled box is $k$-tiled.
 (Follows from $k\!\!=\!\!2$ case by seeing $(k\!\!-\!\!2)$-blocks as tiles.)
\end{lemma}

\paragraph {Proof.} The eight colors of edges fix their location
in 2-frames, forcing open 1-blocks to alternate in the pattern of
2-frames which, by Lemma~\ref{blocks}, extend to 2-blocks. \qed

\begin{cor} Any $2^k\times2^k$ box, extendible to a 3 times wider
 cocentric 1-tiled box, extends to a $(k\!+\!4)$-block.\end{cor}

\paragraph {Induction Basis.} For the simplest enforcement of tiling
decomposition into 1-blocks we can use a 2-periodic \df {parity} bit to
mark \df {odd} lines carrying 0-medians.
 All pointers on odd lines point to odd crossing lines, thus forcing a
period 2 on them.
 One needs only to assure an odd line exists. This can be easily done
with a \df {parity pointer} on even lines, pointing to a crossing odd
line.

\paragraph {\hspace*{1pc} Proof of Corollary.} The box is $k$-tiled
covering the inner box with four blocks sharing a cross. The extension
comes by viewing them as tiles with parity bit reflecting the blocks'
orientation and noting that each cross of $C$ with parity appears in its
open 4-blocks. \qed

 {\bf Proof of Theorem~\ref{E}.} Let $T$ be a $ce$-tiling decomposed,
for each $k$, into $k$-blocks
 with equal frames. Then a shift of $CE$ matches $T$ on all lines of
rank $<k$.
 The shifted $CE$ converge pointwise to $T$, except possibly on their
(unranked) axes.
 By Remark~\ref{R}, the shift increments grow in rank, and so sum to one
2-adic shift.
 Finally, reflections match the brackets on axes. \qed

\subsection {Parsimonious Enforcement of the Grid of 1-Blocks}

First, we reduce the needed parity colors.
 A parity pointer on a single edge suffices,
 so it needs to accompany only one color if we show that $ce$-tilings
cannot skip colors.
 Indeed, all $ce$-crosses are either \df {bends}, \ie have 4 inward
pointers, or \df {passes}, \ie have 1.
 Thus, a third of crosses are bends, up to $O(n)$ accuracy for
 $n\times n$ boxes.
 Moreover, all orientations of bends are equally frequent, alternating
on each line. \qed

Tedious case investigation of~\cite {levitsky} shows $ce$ bits
themselves forcing 1-blocks, rendering parity bits redundant.
A \df {$k$-bar} is a maximal bold or pale segment, $k$ being its length.
 $k>1$ and no bold 3-bar exists since it is easy to see that its middle
link would be connected by a tile to a 1-bar. Levitsky first proves that
each $ce$-tiling has bold 2-bars. Here is a simpler argument for this.

 A third of crosses are bends, so the average bar length is 3.
 Absent bold 2-bars, this average would allow positive density only of
bold 4-bars and pale 2-bars. Tilings with such bars have period 6 and
map onto a $6\times 6$ torus with 3 bold $4\times 4$ squares. But $Z_6$
cannot have three disjoint pairs of points of equal parity! \qed

The rest of \cite {levitsky} analysis assures bold 2-bars two tiles
away at each side of any bold 2-bar. This involves a case-by-case
demonstration that no violation can be centered in a $10\times 10$ box.
The analysis is laborious but may be verifiable by a computer check.

 \section {Acknowledgments}

These remarks were developed in my attempts to understand the classical
constructions of aperiodic tilings while working on~\cite{dls}.
 My main source of information was \cite{ad-bgg} and its explanations
by B. Durand and A. Shen to whom I owe all my knowledge on this topic.


\begin {thebibliography} {99}

 \bibitem[Berger 66]{berger} Robert Berger. The Undecidability
 of the Domino Problem. {\em Mem.Am.Math.Soc.}, {\bf 66}, 1-72, 1966.

 \bibitem[Robinson 71]{robinson} Raphael M.~Robinson.
 Undecidability and Nonperiodicity of Tilings of the Plane.\\
 {\em Inventiones Mathematicae}, {\bf 12}:177--209, 1971.

\bibitem[Myers 74]{myers} Dale Myers. Nonrecursive Tilings of the Plane.
ii. {\em J. Symbolic Logic}, {\bf 39(2)}:286--294, 1974.

 \bibitem[Gurevich Koriakov 72]{gurkor72} Yu.~Gurevich, I.~Koriakov.
 A Remark on Berger's Paper on the Domino Problem.
 {\em Siberian Math. J.}, {\bf 13}:459--463, 1972.

 \bibitem[Allauzen Durand 96]{ad-bgg} Cyril Allauzen, Bruno Durand.
Appendix~A: Tiling Problems. pp.~407--420.\\ In Egon Boerger, Erich Graedel,
Yuri Gurevich. {\em The Classical Decision Problem.} Springer-Verlag, 1996.

\bibitem[Levitsky 04]{levitsky} Alexander Levitsky.
{\em Aperiodicheskie Zamoshcheniya s Malym Alfavitom\\ (= Aperiodic Tilings
 with Small Alphabet)}. In Russian. Master Thesis. Moscow University, 2004.

\bibitem[Durand, Levin, Shen 01]{dls}
 Bruno Durand, Leonid A. Levin, Alexander Shen. Complex Tilings.
 {\em Proc. 33rd Annual ACM Symp. on Theory of Computing},
 July 6-8, pp. 732-739. Heraklion, Crete, Greece, 2001.

 \bibitem[Gacs 01]{gacs} Peter Gacs.
 Reliable Cellular Automata with Self-Organization.
 {\em J.Stat.Phys.},\\ {\bf 103(1,2)}:45--267, 2001.
 Also available at: http://www.arxiv.org/abs/math.PR/0003117 .

 \end{thebibliography}
 \end{document}